\documentclass[journal=jacsat,manuscript=article]{achemso}

\usepackage[version=3]{mhchem} \usepackage{soul}
\usepackage{xcolor}
\usepackage{setspace}
\usepackage{hyperref}

\hypersetup{
    colorlinks=true,
    linkcolor=blue,
    filecolor=magenta,      
    pdftitle={arXiv},
    urlcolor=cyan,
    pdfpagemode=FullScreen,
    }

\makeatletter
\let\l@addto@macro\relax
\makeatother
\usepackage[fontsize=10pt]{scrextend}

\usepackage{graphicx}%
\usepackage{multirow}%
\usepackage{amsmath,amssymb,amsfonts}%
\usepackage{amsthm}%
\usepackage{mathrsfs}%
\usepackage[title]{appendix}%
\usepackage{xcolor}%
\usepackage{textcomp}%
\usepackage{manyfoot}%
\usepackage{booktabs}%
\usepackage{algorithm}%
\usepackage{algorithmicx}%
\usepackage{algpseudocode}%
\usepackage{listings}%

\author{Kseniia Lezhennikova}
\affiliation[Aix Marseille Univ, CNRS, Centrale Marseille, Institut Fresnel,
Institut Marseille Imaging, AMUTech, 13013 Marseille, France]{Aix Marseille Univ, CNRS, Centrale Marseille, Institut Fresnel,
Institut Marseille Imaging, AMUTech, 13013 Marseille, France}
\alsoaffiliation[Multiwave Technologies AG, 3 Chemin du Pre Fleuri, Geneva, 1228, Switzerland]
{Multiwave Technologies AG, 3 Chemin du Pre Fleuri, Geneva, 1228, Switzerland}
\alsoaffiliation[Institute of Photonics and Optical Science (IPOS), School of Physics, The University of Sydney, Sydney, NSW 2006, Australia]
{Institute of Photonics and Optical Science (IPOS), School of Physics, The University of Sydney, Sydney, NSW 2006, Australia}
\email{kseniia.lezhennikova@fresnel.fr}
\author{Sahand Mahmoodian}
\affiliation[Institute of Photonics and Optical Science (IPOS), School of Physics, The University of Sydney, Sydney, NSW 2006, Australia]
{Institute of Photonics and Optical Science (IPOS), School of Physics, The University of Sydney, Sydney, NSW 2006, Australia}
\alsoaffiliation[The University of Sydney Nano Institute, The University of Sydney, Sydney, NSW 2006, Australia]
{The University of Sydney Nano Institute, The University of Sydney, Sydney, NSW 2006, Australia}
\author{Boris T. Kuhlmey}
\affiliation[Institute of Photonics and Optical Science (IPOS), School of Physics, The University of Sydney, Sydney, NSW 2006, Australia]
{Institute of Photonics and Optical Science (IPOS), School of Physics, The University of Sydney, Sydney, NSW 2006, Australia}
\alsoaffiliation[The University of Sydney Nano Institute, The University of Sydney, Sydney, NSW 2006, Australia]
{The University of Sydney Nano Institute, The University of Sydney, Sydney, NSW 2006, Australia}
\author{Redha Abdeddaim}
\affiliation[Aix Marseille Univ, CNRS, Centrale Marseille, Institut Fresnel,
Institut Marseille Imaging, AMUTech, 13013 Marseille, France]{Aix Marseille Univ, CNRS, Centrale Marseille, Institut Fresnel,
Institut Marseille Imaging, AMUTech, 13013 Marseille, France}
\author{Stefan Enoch}
\affiliation[Aix Marseille Univ, CNRS, Centrale Marseille, Institut Fresnel,
Institut Marseille Imaging, AMUTech, 13013 Marseille, France]{Aix Marseille Univ, CNRS, Centrale Marseille, Institut Fresnel,
Institut Marseille Imaging, AMUTech, 13013 Marseille, France}
\author{C. Martijn de Sterke}
\affiliation[Institute of Photonics and Optical Science (IPOS), School of Physics, The University of Sydney, Sydney, NSW 2006, Australia]
{Institute of Photonics and Optical Science (IPOS), School of Physics, The University of Sydney, Sydney, NSW 2006, Australia}
\alsoaffiliation[The University of Sydney Nano Institute, The University of Sydney, Sydney, NSW 2006, Australia]
{The University of Sydney Nano Institute, The University of Sydney, Sydney, NSW 2006, Australia}
\author{Alessandro Tuniz}
\affiliation[Institute of Photonics and Optical Science (IPOS), School of Physics, The University of Sydney, Sydney, NSW 2006, Australia]
{Institute of Photonics and Optical Science (IPOS), School of Physics, The University of Sydney, Sydney, NSW 2006, Australia}
\alsoaffiliation[The University of Sydney Nano Institute, The University of Sydney, Sydney, NSW 2006, Australia]
{The University of Sydney Nano Institute, The University of Sydney, Sydney, NSW 2006, Australia}
\email{alessandro.tuniz@sydney.edu.au}

\title[Article Title]{Broadband terahertz near-field excitation and detection of silicon photonic crystal modes}

\begin{document}

\maketitle

\clearpage

\begin{abstract}
Chip-based terahertz (THz) devices are emerging as versatile tools for manipulating mm-wave frequencies in the context of integrated high-speed communication technologies for potential sixth-generation (6G) wireless applications. The characterization of THz devices is typically performed using far-field techniques that provide limited information about the underlying physical mechanisms producing them. As the library of chip-based functionalities expands, e.g., for tailoring the emission and directional propagation properties of THz antennas and waveguides, novel characterization techniques will likely be beneficial for observing subtle effects that are sensitive to a device's structural parameters. Here we present near-field measurements showing the emission properties of a broadband THz emitter placed in the vicinity of a photonic crystal (PHC) slab. These experiments reveal long-predicted emission properties, but which to our knowledge have yet to be experimentally observed at THz frequencies. We demonstrate three distinct effects between 0.3-0.5\,THz: (i) field suppression at frequencies corresponding to quasi-TE bandgaps (ii) a frequency-dependent directed emission along two distinct pathways for two neighboring frequencies, resulting in a local field concentration; (iii) a re-direction of the emission, achieved by rotating the PHC with respect to the dipole orientation. Simulations reveal that the observed behavior can be predicted from the underlying band structure. Our results highlight the opportunities that PHCs can potentially provide for alignment-free, chip-based 6G technologies. Our experimental technique extends the applicability realms of THz spectroscopy and will find use for characterizing the THz modes supported by true samples, whose inherent imperfections cannot realistically be accounted for by simulations, particularly in highly dispersive frequency bands. \end{abstract}

\clearpage 

\section{Introduction}

The terahertz (THz) frequency range~\cite{jepsen2011terahertz}, which spans 0.1-10\,THz and corresponds to wavelengths between 3\,mm--30\,$\mu$m, holds potential for a wide range of applications including bio-sensing~\cite{markelz2022perspective, smolyanskaya2018terahertz}, non-destructive imaging~
\cite{lee2022frontiers}, spectroscopy~\cite{Seo2022}, space communication~\cite{siegel2007thz}, and telecommunications ~\cite{kleine2011review}. Recent years, in particular, have witnessed a surge in interest in terahertz technologies that may serve the imminent rollout of 6G telecommunication networks with terabits/s data rates~\cite{yang2020navigating},  which will be essential for providing a reliable communication backbone for 
several emerging technologies and systems~\cite{mamaghani2022aerial}, such as interconnected vehicles, augmented and virtual reality, and artificial intelligence. As a result, much effort has been directed towards the development of increasingly efficient, compact, and reliable terahertz interconnect solutions~\cite{headland_gratingless_2021, tan_terahertz_2023}.

In the context of free-space communication, terahertz radiation uniquely provides a bridge between optics and electronics
~\cite{pang2022bridging}, sharing the advantages of both IR waves (high directivity, large bandwidth) and microwaves (reduced diffraction). However, the strong dependence on atmospheric conditions~\cite{ma2018invited}, the large absorption of water and vapour~\cite{federici2016review}, signal interference~\cite{shrestha2022jamming}, the loss of the
long-range free space path~\cite{shams2017photonics}, and the need for precise alignment between the source and receiver~\cite{piesiewicz2008performance}, limit the practicality of THz wireless links. To address this, recent years have been marked by the emergence of waveguide-based (sometimes referred to as \emph{wired}) terahertz communication platforms~\cite{xu2022wired}, which have the potential to overcome some of the above challenges. Examples of short-haul wired terahertz links~\cite{koala_nanophotonics-inspired_2022} ($<5$\,cm) include photonic crystal~\cite{tsuruda2015extremely} (PC), topological~\cite{yang2020terahertz}, and suspended~\cite{headland2020unclad} waveguides; medium-haul links
($\sim 5 \,{\rm cm} - 5\,{\rm m}$) are typically achieved with THz fibers~\cite{islam2020terahertz, nielsen2009bendable}.
Longer-haul links are less common, due to the intrinsic large losses of solid materials relative to other frequency bands, and due to large bend losses in the case of large-area hollow core waveguides~\cite{stefani2021bend}.

The emergence of wired devices has, in turn, been accompanied by a wide range of passive components that need to seamlessly interface between multiple functional devices, including, for example, directional couplers and splitters~\cite{weidenbach20163d}, and frequency multiplexers~\cite{yata2016photonic, cao2022add}, which ideally occupy the smallest possible footprint, and are thus best suited for integration with short-haul devices. The most popular material for short-haul applications is float zone silicon~\cite{dai2004terahertz}, because it exhibits extremely low loss in the THz range, strong confinement (refractive index: 3.42), and can be fabricated using established deep-etching fabrication processes~\cite{bruckner2009broadband}. Most existing designs for beam steering demand appropriately engineered and at times rather sophisticated waveguide arrangements (e.g., using leaky~\cite{headland2021gratingless} and coupled metallic wire~\cite{cao2022add} arrangement, or appropriate defects within photonic crystals~\cite{yata2016photonic} and metasurfaces~\cite{cong2018all, liu2021multifunctional}.) 

On the other hand, it is well known that \emph{bulk} photonic crystals (PCs)~\cite{yu2019two} and metasurfaces~\cite{yermakov2018experimental} exhibit a rich underlying bandstructure which, even without defects, can enhance or suppress emission of nearby sources~\cite{noda2007spontaneous}, e.g. dipole antennae~\cite{hoeppe2012direct} (in the microwave regime) and quantum dots~\cite{kounoike2005investigation} (in the near-IR), or provide diffractionless propagation~\cite{iliew2004diffractionless} or long-range supercollimation~\cite{kosaka1999self, rakich_achieving_2006}, enabling alignment-free self-guiding~\cite{chigrin2003self, yu2003bends} at specific frequencies. Such effects rely on a tailored local density of states via a flattening of the isofrequency contours at specific frequencies~\cite{gralak_anomalous_2000, matthews_self-collimation_2007, zhaolin_lu_experimental_2005, lu_experimental_2006}. In such regions, self-guiding~\cite{yu2003bends} or directed emission~\cite{chigrin2003self, yu2019two} are supported, depending on whether the input is a collimated beam or a point dipole. Although these effects have been known for more than 20 years, self-collimation effects after far-field coupling have, to the best of our knowledge, only been measured directly at microwave and near-IR frequencies \cite{zhaolin_lu_experimental_2005, lu_experimental_2006, combrie2009directive, yermakov_experimental_2018, rakich_achieving_2006, martinez2006generation, le2014simultaneous, rotenberg2014mapping, matthews2009experimental, trull2011formation}. Directed emission from a dipole antenna directly placed the near field has been observed in the microwave range~\cite{bulu2003highly, yarga2008degenerate, yermakov_experimental_2018, bulu2005compact}, with a maximum reported frequency of 40\,GHz in photonic crystals~\cite{zhaolin_lu_experimental_2005}, due to the experimental challenges associated with other spectral regions. At higher frequencies, most experiments either excite or collect radiation from an external source far from the sample, which provides a lot of information about the overall transmission and reflection properties into the far field, but no information on the field profiles within the sample itself, and any subtle physics that emerges therein.

Here we present experiments revealing exquisitely controlled emission of a broadband terahertz dipole emitter and detector located in close proximity to a terahertz photonic crystal formed by a silicon slab containing periodic sub-wavelength holes.
We perform near-field emission and detection experiments showing electric field suppression when the dipole frequency lies within the two distinct quasi-TE bandgaps at 0.35\,THz and 0.5\,THz, as well as local field concentration at frequencies near a band edge. Remarkably, we find that two closely spaced frequency channels near 0.455\, THz ($\pm 2\%$) show directed emission towards distinct and complementary spatial locations, as a result of the underlying photonic crystal bandstructure -- an aspect which has, to the best of our knowledge, not been observed to date at any frequency, presumably because of the narrow bandwidths involved.  Furthermore, we show that such emissions can be redirected along different paths simply by changing the photonic crystal orientation relative to the dipole. The phenomena in this paper may appear to be reminiscent of the superprism effect~\cite{kosaka1998superprism}, so it is worth pointing out some key differences. In superprisms, an external collimated monochromatic beam refracts from the photonic crystal edge along vastly different directions inside it upon small changes in the incident angle. In contrast, here we show that a dipole antenna radiating near the photonic crystal surface results in directed quasi-collimated emission along different directions at two distinct neighboring frequencies. 

Our experiments are in good agreement with simulations, which predict the frequencies at which such effects occur with an error of $3\%$. This highlights the importance of performing full measurements on true samples, including all and any imperfections, which cannot be done with simulations alone. Our results also show the potential of using structured media to achieve tailored emission of 6G antenna networks on chip-based platforms.

\begin{figure}[t!]
    \centering
    \includegraphics[width=\linewidth]{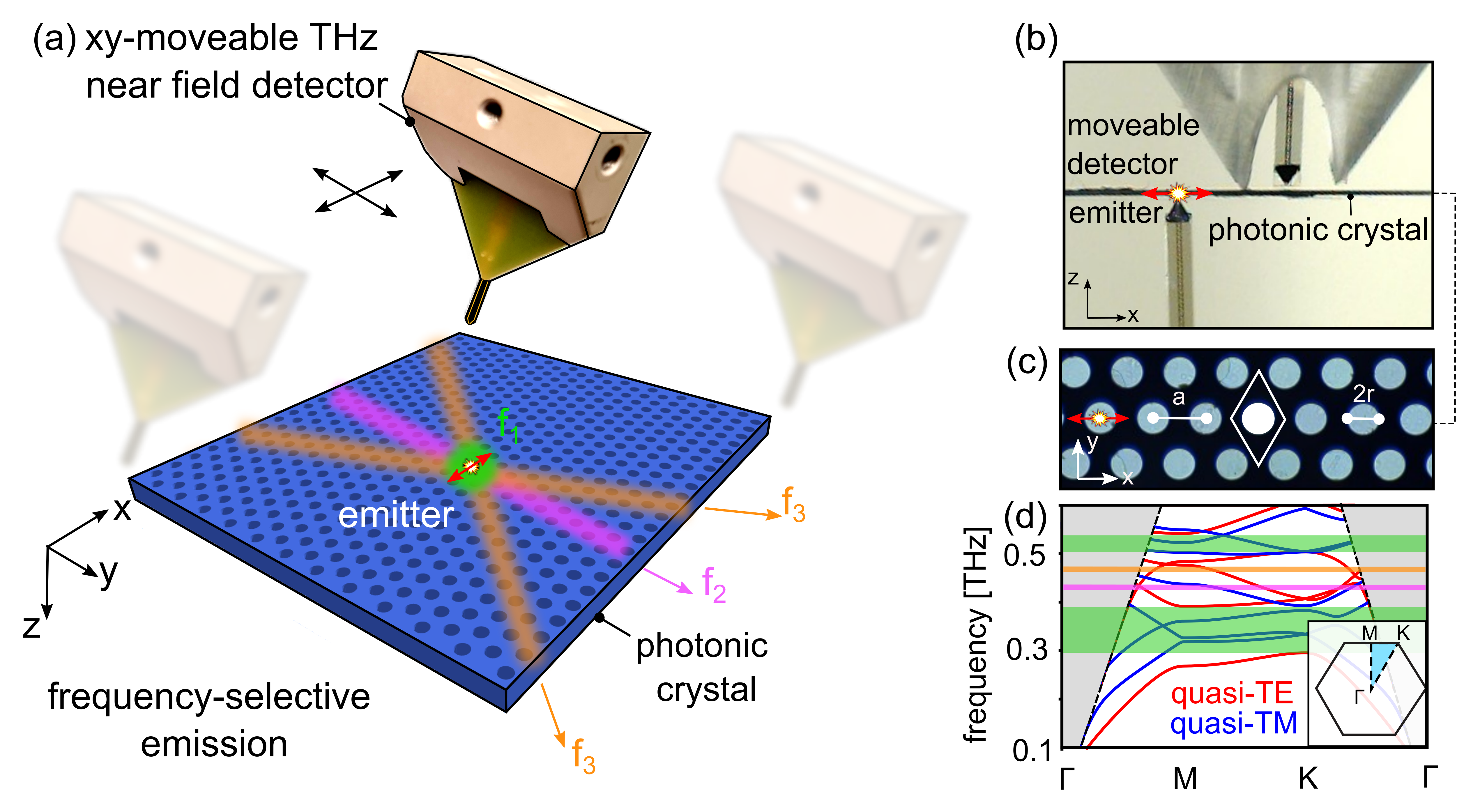}
    \caption{(a) Schematic representation of the experiment. The emission of a broadband terahertz near-field $x$-polarized point dipole near the surface of a silicon photonic crystal membrane subtly depends on frequency, as a result of the underlying bandstructure. At a frequency $f_1$ (green), the dipole emission into the photonic crystal is suppressed. At a different frequency $f_2$ (pink), the dipole emits preferentially in particular directions, leading to local field concentration. At a neighboring frequency $f_3$ (orange), the direction of emission changes.
    Measurements are performed with a moveable terahertz near-field detector on the surface of the sample. (b) Photograph of the associated experimental configuration. The fixed $x$-oriented terahertz dipole emitter (red arrows) and moveable terahertz detector are placed on either side of the photonic crystal slab. (c) Photonic crystal micrograph. Black: silicon; white: air; period: $a=240~\mu{\rm m}$; hole radius: $r=72~\mu{\rm m}$. The unit cell is highlighted in white, the emitter dipole direction is shown by red arrows. (d) Calculated photonic crystal band structure in the first Brillouin zone, shown in the inset. Green-shaded regions highlight the first and second quasi-TE bandgap, respectively, while purple and orange regions highlight the first and second frequency of the directional emission in (a).}
    \label{fig1}
\end{figure}

\section{Results and Discussion}

Figure~\ref{fig1}(a) shows a schematic of our experimental configuration and approach. We consider a photonic crystal membrane composed of a silicon slab containing a hexagonal lattice of subwavelength diameter holes. A horizontally polarized broadband electric dipole (nominal emission frequency range: 0.2-2\,THz) is placed near its surface. The
resulting field pattern at each frequency depends on the underlying band structure, discussed below, and is measured on the surface of the photonic crystal membrane by a terahertz near-field antenna. Figure~\ref{fig1}(b) shows a photograph of the experimental setup, which is a modified terahertz time-domain spectroscopy system~\cite{stefani_flexible_2022}, wherein two near-field $x$-polarized antenna probes~\cite{protemics} are used as the emitter and the detector on opposite sides of the membrane, as described in the Methods section. The electric field is polarized in $x$, using the sample orientation and reference frame shown in Fig.~\ref{fig1}.   
We estimate the distance between the sample and the emitter/detector to be $\sim 200\, \mu{\rm m}$, which reduces the risk of sample and antenna damage during such large area scans. 

Note that a similar setup was recently used to measure the terahertz Local Density of States on metal and dielectric surfaces~\cite{ter2023direct}, the symmetry protection properties of Bound States in the Continuum~\cite{van2021unveiling}, and other near-field topological waveguide effects~\cite{yang2022topology} -- with a notable difference that here, placing the emitter and detector on opposite sides of the sample (as opposed to placing both emitter and detector on the same side~\cite{van2021unveiling}) has two important advantages: (i) it allows us to scan the \emph{entire} sample surface because we are not limited to regions that are far from the bulky emitter antenna; (ii)  by detecting the evanescent
field of the slab modes excited by the emitter itself, it eliminates the possibility of measuring the direct contribution of the emitted field, which is mediated by the slab modes' surface waves. Such an approach has been reported in the microwave range, e.g., experimentally characterizing hyperbolic media~\cite{yermakov_experimental_2018}, topological effects~\cite{yang2018ideal}, and synthetic frequencies~\cite{guan2023overcoming}  where sources and detectors are macroscopic, but has to the best of our knowledge not been shown at THz frequencies.

\begin{figure}[t!]
    \centering
\includegraphics[width=\linewidth]{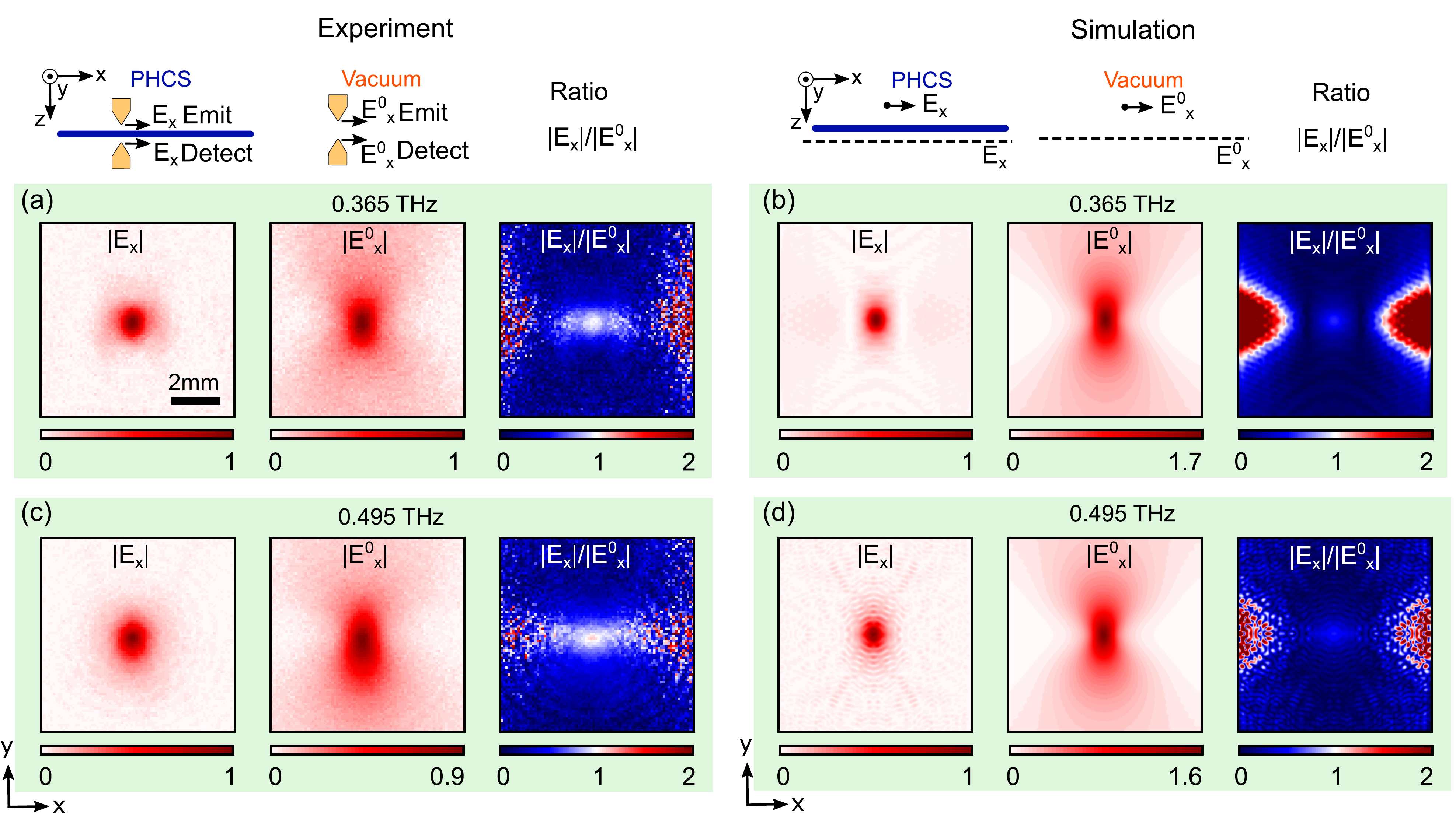}
    \caption{Dipole emission suppression within the two quasi-TE bandgaps. (a) Measured $x$-component of the electric field magnitude $|E_x|$ near the surface of the slab for an $x$-polarized emitter placed on the opposite side, at 0.365\,THz (left). Also shown in $|E_x^0|$ at the same location after removing the slab  (middle), and the ratio $|E_x|/|E_x^0|$  (right), revealing dipole emission suppression in $y$. (b) Corresponding  FDTD numerical simulations, taking an $x$-polarized 
 dipole on one side of the slab as per the top schematics, at 0.35\,THz. (c) and (d) respectively show the corresponding experiment (0.495\,THz) and simulation (0.48\,THz) in the second bandgap.  All window sizes are 8\,mm $\times$ 8\,mm. The associated scale bar, shown only in (a) for clarity, is 2\,mm.}
    \label{fig2}
\end{figure}

Figure~\ref{fig1}(c) shows a micrograph of the photonic crystal, formed by hexagonally arranged air holes in float zone silicon, and fabricated using deep etching~\cite{bruckner2009broadband}. The thickness of the slab is $250\,\mu {\rm m}$, the radius of the hole is $r=72~\mu{\rm m}$, and the pitch from center to center is $a=240~\mu{\rm m}$. 
The photonic crystal band structure, calculated using the plane-wave expansion method~\cite{minkov2020inverse}, is shown in Fig.~\ref{fig1}(d): as expected~\cite{photonic_crystals_book}, this structure has two quasi-TE band gaps at $\sim0.35$ THz and $\sim 0.5$ THz (shaded green), which should suppress the dipole emission. Furthermore, close to the band-edge (shaded in red), we expect directed emission that subtly depends on the relative orientation of the dipole with respect to the photonic crystal lattice~\cite{chigrin2003self, yu2019two}. We now investigate each case in detail.

\subsubsection*{Bandgap emission suppression}

We begin by considering the lowest-frequency bandgap as per Fig.~\ref{fig1}(d). Figure \ref{fig2}(a) (left) shows the measured $x$-polarized electric field amplitude $|E_x|$ detected at 0.365\,THz on the sample surface, as per the top schematic. We notice a central bright region, corresponding to the location of the source on the opposite side of the sample, but no vertical diffraction on either side, suggesting that emission in the $y$-direction has been suppressed. To confirm this, we remove the sample, leaving all other conditions unchanged, and repeat the measurements to obtain $|E_x^0|$. The results are shown in the middle window of Fig.~\ref{fig2}(a), with no emission in each $y$-direction as expected for an $x$-polarized dipole. To quantify the emission suppression in the bandgap, we take the ratio $|E_x|/|E_x^0|$ (right window), which clearly shows a suppression of the emission in the vertical direction. Note that, in displaying this amplitude ratio, the color scale has been chosen so that blue and red are below and above unity, respectively.

A comparison with the corresponding Finite Difference Time Domain simulation (CST Microwave Studio) is shown in Fig.~\ref{fig2}(b). The photonic crystal was modeled as a slab with a thickness of 250\,$\mu$m made of dielectric with a refractive index $n=3.42$~\cite{dai2004terahertz} between 0.3\,--\,0.50\,THz, containing hexagonally arranged holes with diameter and pitch as in our sample specifications, using a slab area of 1\,cm\,$\times$\,1\,cm and open boundary conditions in the $x$- and $y$- directions. The $x$-polarized dipole source and the electric field detector are located at a distance of 200\,µm, on opposite sides of the photonic crystal surface, for fair comparison with experiments. The corresponding $E_x^0$ is obtained by considering the same simulation space without the photonic crystal. We find excellent agreement between the experiment and simulation, confirming the dipole emission suppression in $y$ due to the TE-band gap.

The large bandwidth of our terahertz probe also allows us to probe the emission suppression at the \emph{second} quasi-TE bandgap at 0.495\,THz (see Fig.~\ref{fig2}(c)), half an octave higher in frequency. Figure~\ref{fig2}(c) and (d) show the corresponding experiment and simulations at that frequency, both showing emission suppression, in excellent agreement with each other. These experiments reveal the possibility of achieving a comparable and tailored (here: suppressed) emission profile over distinct and widely separated frequency bands, all using the same sample.

\subsubsection*{Band-edge directed emission and field concentration}

As a next step, we consider the emission profile of two frequencies close to the lower edge of the second quasi-TE bandgap ($\sim$0.45\,THz). It is well known that in these regions self-guiding~\cite{yu2003bends} or directed emission~\cite{chigrin2003self, yu2019two} are supported, depending on whether the source is a collimated beam or a point dipole. Such behavior is a notoriously narrow band~\cite{yu2003bends}, and therefore highly sensitive to any fabrication-induced deviations from the numerical design, thus requiring accurate experimental measurements after fabrication to complement the design phase. Here the large bandwidth of the THz source allows us to scan all frequencies in this region to identify at which frequencies directed emission occurs. 

\begin{figure}[t!]
    \centering
    \hspace*{-1.75cm} 
\includegraphics[width=1.2\linewidth]{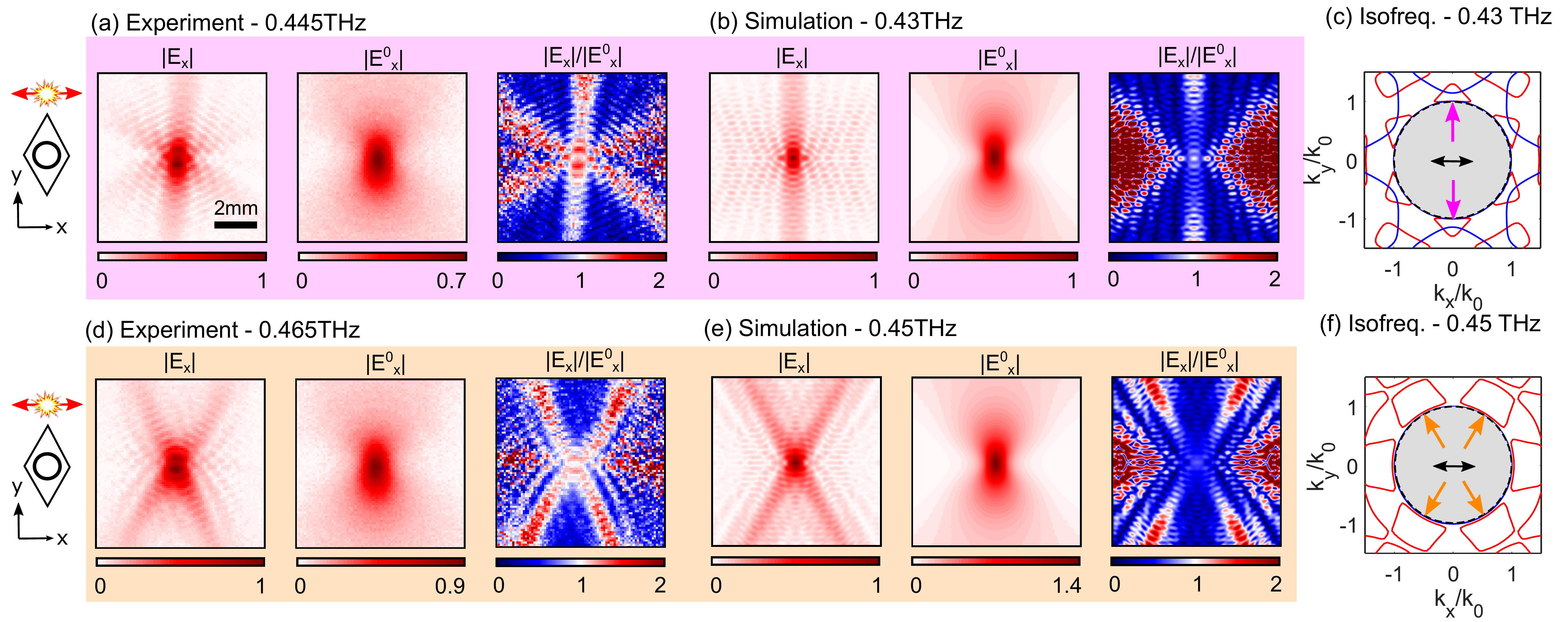}
    \caption{Directed emission of the $x$-polarized dipole. The photonic crystal and dipole are oriented as in Fig.~\ref{fig1}(c), as highlighted by the left-most schematic. (a) Measured $x$-component of the electric field magnitude $|E_x|$ versus position on the surface of the slab for an $x$-polarized emitter placed on the opposite side, at 0.445\,THz (left). Also shown in $|E_x^0|$ at the same location after removing the slab (middle) and the ratio $|E_x|/|E_x^0|$  (right), revealing directed emission in $y$ accompanied by a field concentration. (b) Corresponding FDTD numerical simulations, taking a $x$-polarized dipole on one side of the slab as per the top schematics in Fig.~\ref{fig2}. (c) Corresponding isofrequency contour for quasi-TE (red) and quasi-TM (blue) modes. The black arrow corresponds to dipole orientation, which preferentially excites the quasi-TE Bloch modes that are close to the light line and propagated perpendicular to it (purple arrows). (d) and (e) respectively show the corresponding experiment (0.465\,THz) and simulation (0.45\,THz), and (f) isofrequency contour. Note that the Bloch modes close to the light line which are preferentially excited by the dipole here lie on the diagonal axes, as observed in both experiment and simulation. The photonic crystal orientation is as shown in Fig.~\ref{fig1}(c). The window sizes in (a), (b), (d), and (e) are 8\,mm $\times$ 8\,mm. The associated scale bar, shown only in (a) for clarity, is 2\,mm. All the field-maps are normalized by the maximum of $|E_x|$.}
    \label{fig3}
\end{figure}

Figure~\ref{fig3} shows the results of the same experiments and simulations shown in Fig.~\ref{fig2} but in a different spectral region. Figure~\ref{fig3}(a) (left) shows the measured $x$-polarized electric field amplitude $|E_x|$ detected at 0.445\,THz on the sample surface, for a PC unit cell oriented as shown on the left-most column, here showing strong vertical directed emission and weak emission in two diagonal regions. Taking the ratio with respect to the dipole intensity pattern with the sample removed (Fig.~\ref{fig3}(b), middle) clearly shows a field concentration along the vertical and diagonal regions (Fig.~\ref{fig3}(b), middle). Note in particular that the field is concentrated by a factor of two along the vertical region (i.e., $|E_x|/|E_x^0|>1$), as opposed to mere spatial filtering. This corresponds to an increase of 6\,dB in field intensity, despite the presence of the silicon. Simulations indicate that, compared to a bare silicon slab, the relative field concentration in the photonic crystal slab is even greater -- of order 10-30 times -- as illustrated in Figure S1 of the Supporting Information.

These measurements are in good agreement with simulations at 0.43\,THz, which is 0.015\,THz from the measured frequency, most likely due to the extreme sensitivity of these effects to any small deviations in the sample from the ideal design (e.g., imperfections in the fabrication process, difference in hole size, pitch, slab thickness, and refractive index). The associated frequency contours for the quasi-TE modes (red) and quasi-TM modes (blue), shown in Fig.~\ref{fig3}(c), allow us to interpret the observed behavior: a horizontally polarized dipole (black arrow) excites quasi-TE Bloch modes with a nearly flat contour close to the light line at $(k_x/k_0,k_y/k_0) \sim (0,\pm 1)$, corresponding to vertical emission. Note, the diagonal quasi-TE modes near the light line (black dashed line) at $(k_x/k_0,k_y/k_0) \sim (\pm \sqrt{3}/2,\pm 1/2)$ are also excited, albeit more weakly, due to a smaller overlap between the point dipole field pattern and the diagonally-directed Bloch modes. Our analysis is consistent with simulations that consider the 2D Fourier transforms of the measured complex fields, which overlap well with the simulated isofrequency contours, as shown Figure S2 of the Supporting Information.

Finally, we consider field amplitude measurements at a slightly higher frequency (0.465\,THz), shown in Fig.~\ref{fig3}(d). In this case, we observe emission along two diagonal axes, now forming an ``X''-shaped pattern (left), i.e., the emission direction is rotated by $\pi/6$ with respect to the case in Fig.~\ref{fig3}(a). After considering the bare dipole (middle), the associated ratio (right) reveals directed emission and field concentration (as opposed to mere filtering) along the diagonal direction (red regions), in good agreement with the corresponding simulations at 0.45\,THz shown in Fig.~\ref{fig3}(e). The associated calculated isofrequency contour, shown in Fig.~\ref{fig3}(f) now reveals that there are only quasi-TE Bloch modes (red line) that are close to the light line, and which are rotated by $\pi/6$ with respect to those shown in Fig.~\ref{fig3}(c) and highlighted by the orange arrows pointing to $(k_x/k_0,k_y/k_0) \sim (\pm 1/2,\pm \sqrt{3}/2)$. Note that in this case, the overlap integral between the $x$-polarized dipole and the quasi-TE Bloch modes at $(k_x/k_0,k_y/k_0) \sim (\pm 1,0)$ is zero so that no field is emitted along $x$. 

\begin{figure}[t!]
    \centering
    \includegraphics[width=0.8\linewidth]{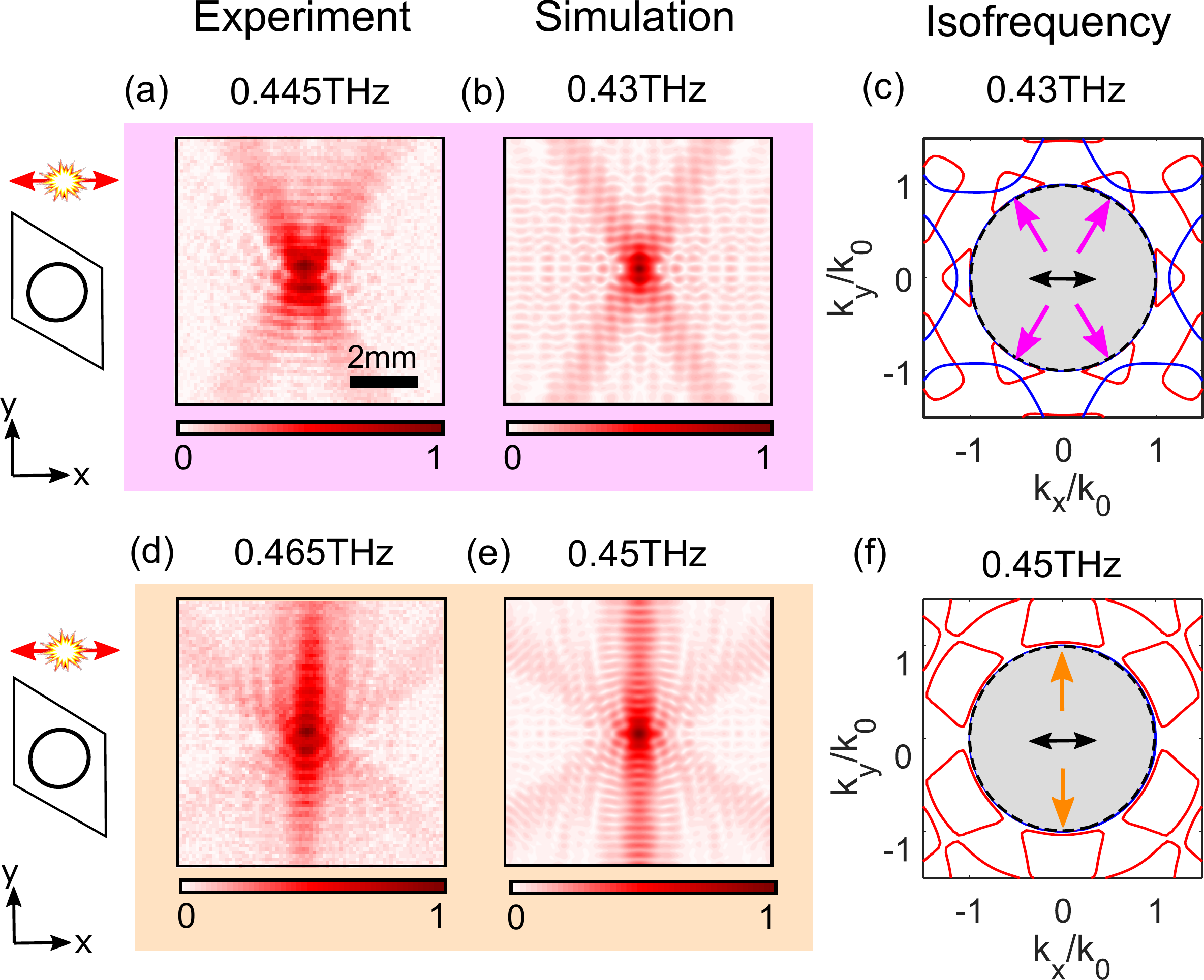}
    \caption{Redirection of the emission direction by rotating the photonic crystal with respect to the dipole, as highlighted by the left-most schematics (a) 
    Measured $x$-component of the electric field magnitude on the surface of the slab for an $x$-polarized emitter (left) and a $y$-polarized emitter (right) placed on the opposite side and at 0.445\,THz, where crystal orientation is as shown in Fig.~\ref{fig1}(c). (b) Corresponding numerical simulations at 0.43\,THz. (c) and (d) show the corresponding experiment and simulation at 0.465\,THz, and 0.45\,THz. Black arrows in (c),(f) indicate dipole orientation on the surface. The spatial scale bar, shown only in (a) for clarity, is 2\,mm.}
    \label{fig4}
\end{figure}

The above experiments and analysis immediately suggest a pathway for redirecting the field emitted by the point dipole, simply by rotating the photonic crystal~\cite{yu2019two}. Figure~\ref{fig4}(a) (left) shows the measured $x$-polarized electric field amplitude, $E_x$, detected at a 0.445\,THz on the sample surface, when the photonic crystal unit cell is rotated by $\pi/6$ with respect to Fig.~\ref{fig3}, as shown by the left-most schematics of Fig.~\ref{fig4}. In this configuration, and in contrast to the results shown in Fig.~\ref{fig3}, we observe an ``X'' shaped emission pattern at this frequency, in agreement with the simulations at 0.43\,THz shown in Fig.~\ref{fig4}(b). This result is an immediate consequence of the associated rotation of the isofrequency contours, as shown in Fig.~\ref{fig4}(c). As a result, the slightly higher frequency of 0.465\,THz shows a predominantly $y$-directed emission (Fig.~\ref{fig4}(d)), in agreement with simulations (Fig.~\ref{fig4}(e)), because of the associated rotated isofrequency contour (Fig.~\ref{fig4}(f)). This clearly indicates that we can control the emission direction not only by changing the excitation frequency but also by altering the polarization of the source dipole with respect to the photonic crystal orientation, providing potential additional avenues for both space- and frequency-division multiplexing on a single chip-based platform.

Note that in our experimental configuration, the emitter and detector antenna are placed at a distance $D$ from the edge of the photonic crystal such that it lies in the radiating near-field (i.e., such that $<\lambda/2\pi<D<\lambda$~\cite{yaghjian1986overview}). At this position, the higher-$k$ Bloch modes cannot be accessed, as shown by additional simulations in Fig.~S2 of the Supporting 
Information, which also considers the 2D Fourier transforms of the measured complex fields. The higher-$k$ Bloch modes can be accessed by placing appropriately polarized dipoles \emph{inside} the photonic crystal, as shown in Fig.~S2 of the Supporting Information, but are inaccessible under the current experimental configurations, and which would, in any case, excite unwanted high-$k$ modes and thus be detrimental to directional emission in the present context. Finally, our simulations suggest that the salient features of the measured images are only weakly dependent on the detector distance, provided that it is within approximately one wavelength of the sample surface. The dependence of the field pattern with respect to the detector's $z$-position is shown in Fig. S3 in the Supporting Information. 

\section{Conclusion}

Photonic crystal membranes are a powerful tool to suppress, concentrate, and redirect the field emitted by terahertz antennas when they are placed close to their surface. To demonstrate this, we developed a novel terahertz near-field emission/detection setup that relies on placing each antenna on opposite sides of the sample. Our approach is generally suited to characterize the subtle and sensitive directional emission properties of any planar geometry. In this particular case, our near-field measurements revealed a reduction in the emitted field inside two bandgaps at 0.35\,THz and 0.5\,THz. Additionally, we observed and explained a previously unreported phenomenon of directional emission and field concentration along either vertical- and diagonal- directions for two frequency channels near 0.455\,THz, spaced only 20\,GHz apart. The signals were then re-oriented simply by rotating the relative axis of the dipole orientation and underlying photonic crystal. Our results show that bulk photonic crystals provide a potential pathway for simple short-haul 6G circuit elements 
without needing any additional defects or materials, and which could, for example, be directly connectorized to suspended silicon waveguides. Future designs might consider integrating several terahertz dipole antennas on photonic crystal surfaces to direct distinct polarizations along separate spatial channels at different frequencies. Connecting these channels to waveguides at the edge of the chip would then form a monolithic frequency-dependent power splitter -- Fig. S4 in the Supporting Information shows a numerical example illustrating its operation using an analogous two-dimensional photonic crystal connected to surrounding step index waveguides. Our approach uses the frequency-dependent photonic crystal band structure, which provides a much richer parameter space than simple waveguides, whose geometry dictates the direction of light propagation at all frequencies. Changing the hole size and spacing throughout the chip could provide additional flexibility with regards to the accessible frequencies for directed emission, within the symmetry constraints of the photonic crystal. Related application examples that warrant revisiting and detailed near field experimental characterization at terahertz frequencies include wavelength-dependent self-guidance~\cite{turduev2013extraordinary}, polarization dependent splitting~\cite{yasa2017polarization} and multiplexing~\cite{ito2018wavelength}.

Our experiments suggest that wider classes of photonic structures, for example, square photonic crystals~\cite{chigrin2003self}, topologically protected edge states~\cite{yang2020terahertz}, Dirac points~\cite{bittner2010observation}, or hyperbolic media~\cite{tuniz2013metamaterial}, likely provide many opportunities for controlling the emission of terahertz antennas in unexpected ways. Due to the extreme sensitivity of photonic crystal band structures to specific designs, our experimental technique will provide a valuable tool for refining device development due to practical limits during fabrication. In the field of quantum optics, strongly modified emission properties induced by the photonic crystal might also be used to mediate the strong anisotropic dipole--dipole coupling of individual quantum emitters~\cite{yu2019two}, the development of which at terahertz frequencies is in a nascent stage~\cite{kibis2009matter, groiseau2023single_test}. More broadly, this terahertz near-field excitation and detection scheme -- which uniquely measures evanescent fields mediated by slab modes over the entire sample surface -- may provide a useful and versatile tool for better understanding or revealing new physics over broad bandwidths, particularly in situations where large-scale numerical simulations become prohibitively slow and computationally demanding.

\section{Methods}

\subsection{Terahertz near field emission and detection}

We use a commercially available THz Time Domain Spectroscopy (THz-TDS) System (Menlo TERAK15), which relies on THz emission from biased photoconductive antennas (Protemics) that are pumped by fiber-coupled near-infrared pulses (pulse width: 90\,fs; wavelength: 1560\,nm) in combination with a second harmonic generaion (SHG) module (Protemics) which converts the laser to 780\,nm to excite the low temperature GaAs photocurrent needed for terahertz generation. The field emitted by the terahertz antenna couples to the Bloch modes of the photonic crystal. On the other side of the sample, the resulting field is sampled as a function of the time delay via an infrared probe pulse on another identical near-field photoconductive antenna module, placed on a movable stage, which forms the THz detector. The near-field is thus spatio-temporally resolved at every point via a raster scan (step size: $100\,\mu{\rm m}$; window size: $8 \times 8\, \text{mm}$) with sub-THz cycle temporal resolution, so that Fast Fourier transforms of the temporal response at each pixel position provide the spectral information, including amplitude and phase.

\subsection*{Acknowledgments} 

This work is funded in part by the Australian Research Council Discovery Early Career Researcher Award (DE200101041) and Future Fellowship (FT200100844). This work was performed in part at the NSW node of the Australian Nanofabrication Facility. This research was conducted within the context of the International Associated Laboratory ``ALPhFA: Associated Laboratory for Photonics between France and Australia''.  We gratefully acknowledge Gloria Qiu and Jackie He from the University of Sydney Research and Prototype Foundry for assistance with sample fabrication, and Alex Y. Song for fruitful discussions. 

\providecommand{\latin}[1]{#1}
\makeatletter
\providecommand{\doi}
  {\begingroup\let\do\@makeother\dospecials
  \catcode`\{=1 \catcode`\}=2 \doi@aux}
\providecommand{\doi@aux}[1]{\endgroup\texttt{#1}}
\makeatother
\providecommand*\mcitethebibliography{\thebibliography}
\csname @ifundefined\endcsname{endmcitethebibliography}
  {\let\endmcitethebibliography\endthebibliography}{}

\clearpage

\renewcommand{\thefigure}{S\arabic{figure}}
\setcounter{figure}{0}

\section*{Supporting Information}

\begin{figure}[h!]
    \centering    \includegraphics[width=0.8\linewidth]{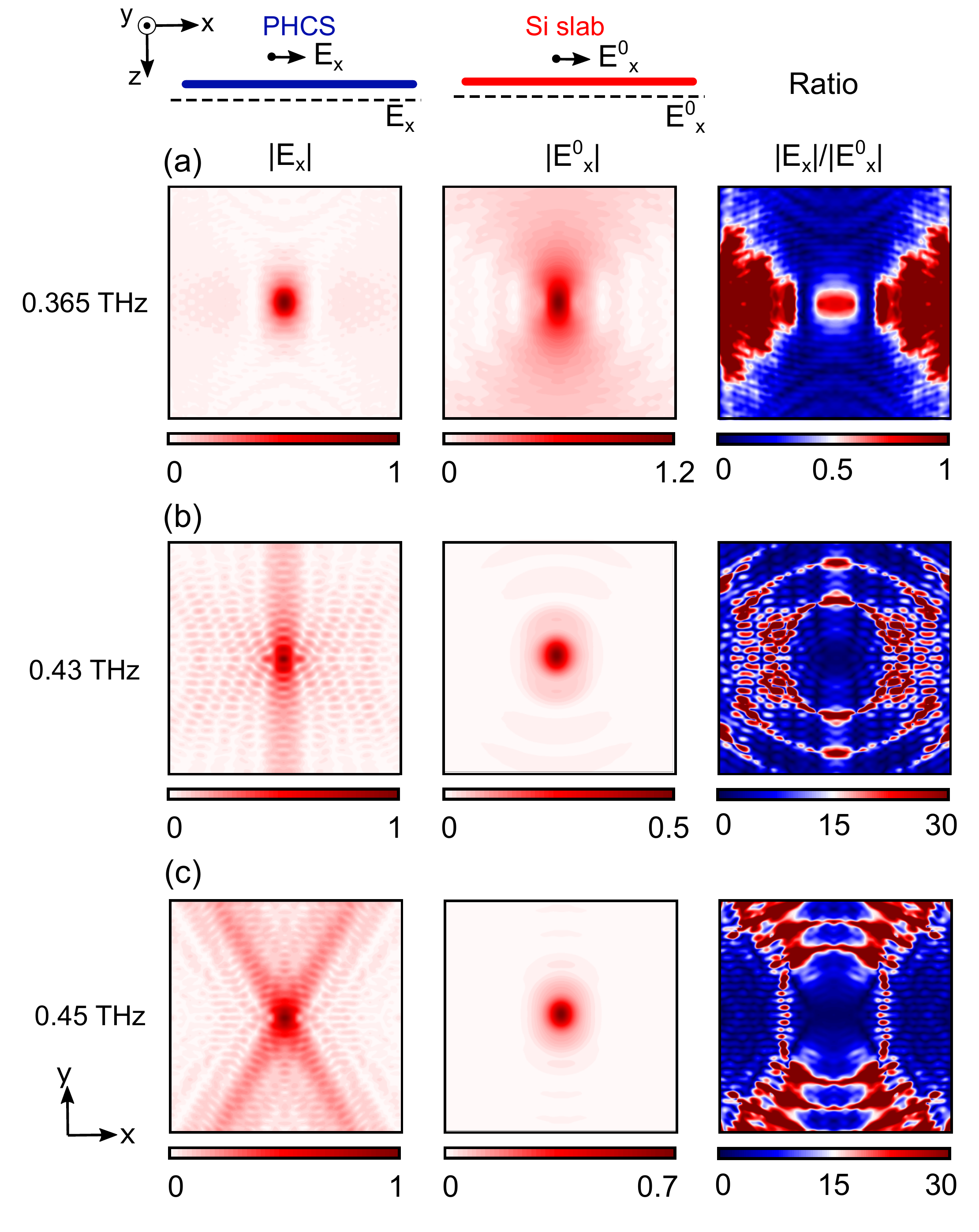}
    \caption{Comparison of the field suppression and concentration in the photonic crystal slab with respect to an unstructured silicon slab for an $x$-polarized dipole. All simulation parameters are analogous to those of Fig. 2(a) in the main manuscript. Simulated $x$-component of the electric field magnitude versus position at 0.365\,THz on the surface of the photonic crystal slab represented by a blue line ($|E_x|$, left), on a bare silicon slab (represented by a red line) at the same position ($|E^0_x|$, middle), and their ratio (right). (b) and (c) show the corresponding simulations at 0.43\,THz and 0.45\,THz respectively, revealing a $>$10-30 fold field concentration along the vertical- and diagonal- directions, even greater than free space. This is largely explained by the proximity of photonic crystal Bloch modes to the light line, compared to the guided modes of unstructured slabs. Window sizes are 8\,mm $\times$ 8\,mm. The field-maps in the first two columns are normalized by the maximum of $|E_x|$ at each frequency. }
    \label{figS2}
\end{figure}

\begin{figure}[h!]
    \centering    \includegraphics[width=\linewidth]{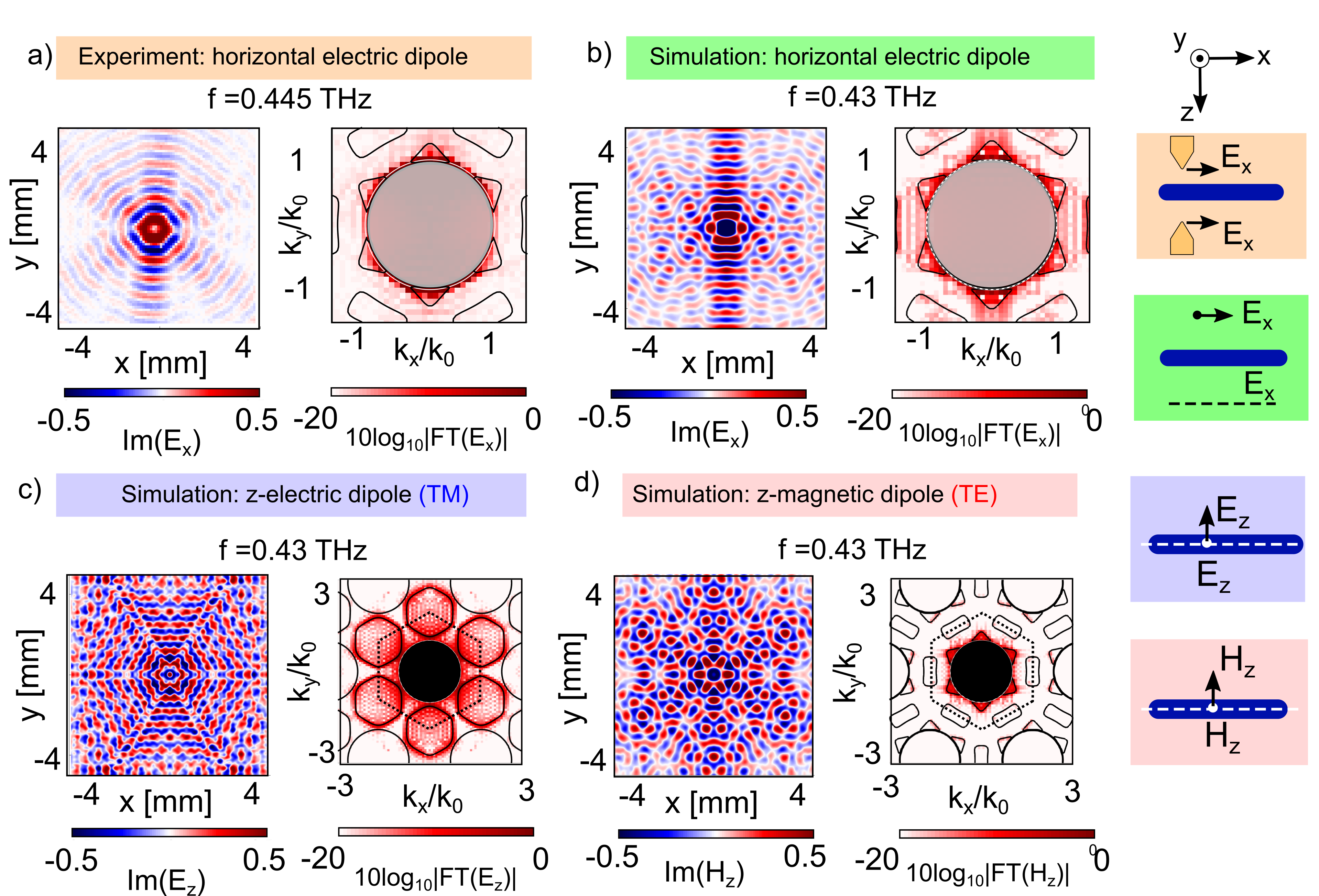}
    \caption{Additional analysis of the experiments and simulations demonstrating the emission properties of the point dipole near a photonic crystal. (a) Measurements of the imaginary part of the transmitted electric field ($x$-component) of an $x$-polarized electric dipole in near-contact with the photonic crystal slab (PHCS) (left), associated amplitude of the 2D Fourier transform (FT) of the complex field (right) at 0.445\,THz and on a dB scale, superimposed with simulated TE isofrequency contours. Its regions of maximum intensity overlap with the part of the isofrequency contours that are close to the light line. (b) Corresponding numerical simulations at 0.43\,THz, showing good overall agreement. The dipole excitation and field detection are at 200\,$\mu {\rm m}$ from the edge of the membrane. (c) and (d) respectively show the corresponding simulations when a $z$-polarized electric- or magnetic- dipole is placed at the center of the photonic crystal. In this case, the high-$k$ quasi-TM and quasi-TE Bloch modes can be excited independently. In contrast, the $x$-polarized dipole in b) only excites a subset of the quasi-TE modes shown in d). The data set in (a) and (b) is the same as Fig.~3 in the main manuscript. Schematics on the right, showing reference frame and experimental/simulation configurations, are color-coded as per (a)--(d). The dark blue region represents the photonic crystal.}
    \label{figS1}
\end{figure}

\begin{figure}[h!]
    \centering  \includegraphics[width=0.69\linewidth]{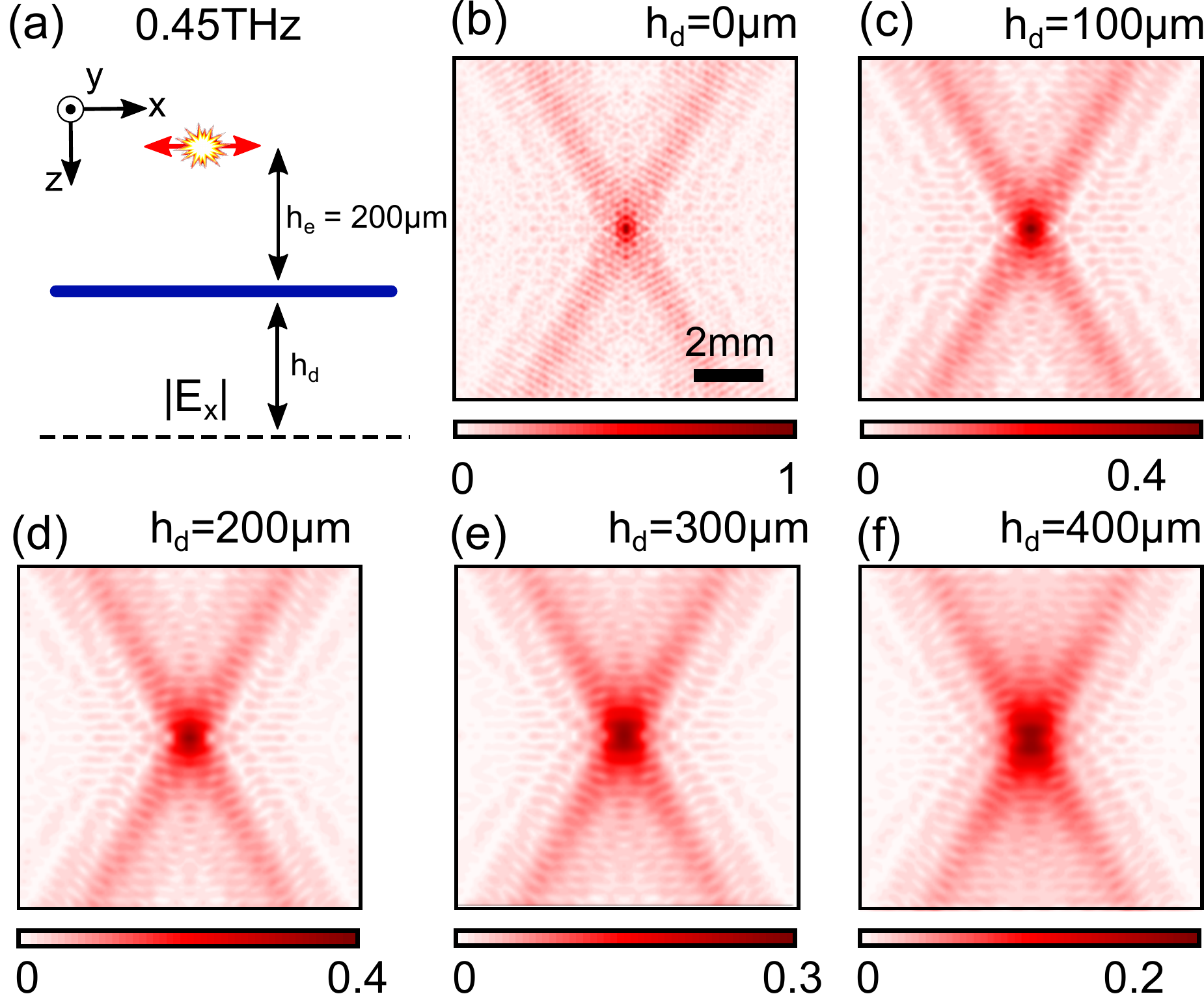}
    \caption{Additional FDTD simulations showing the effect of increasing detector distance. (a) We consider the same emitter configuration as Fig. 3(e) of the main manuscript at 0.45\,THz, keeping the distance $h_e =200\,\mu{\rm m}$ between emitter and photonic crystal constant, but vary the distance $h_d$ between detector and photonic crystal edge. The field magnitude $|E_x|$ is shown at (b) $h_d=0\,\mu {\rm m}$, (c) $h_d=100\,\mu {\rm m}$, (d) $h_d=200\,\mu {\rm m}$ (e) $h_d=300\,\mu {\rm m}$, (f) $h_d=400\,\mu {\rm m}$. All amplitudes are normalized with respect to the maximum in (b). Note that the overall emission pattern is nominally unchanged, but the field amplitude goes down as the detector distance increases. By reciprocity, the same is true when keeping $h_d$ constant and changing $h_e$.}
    \label{fig_revised}
\end{figure}

\begin{figure}[h!]
    \centering    \includegraphics[width=0.69\linewidth]{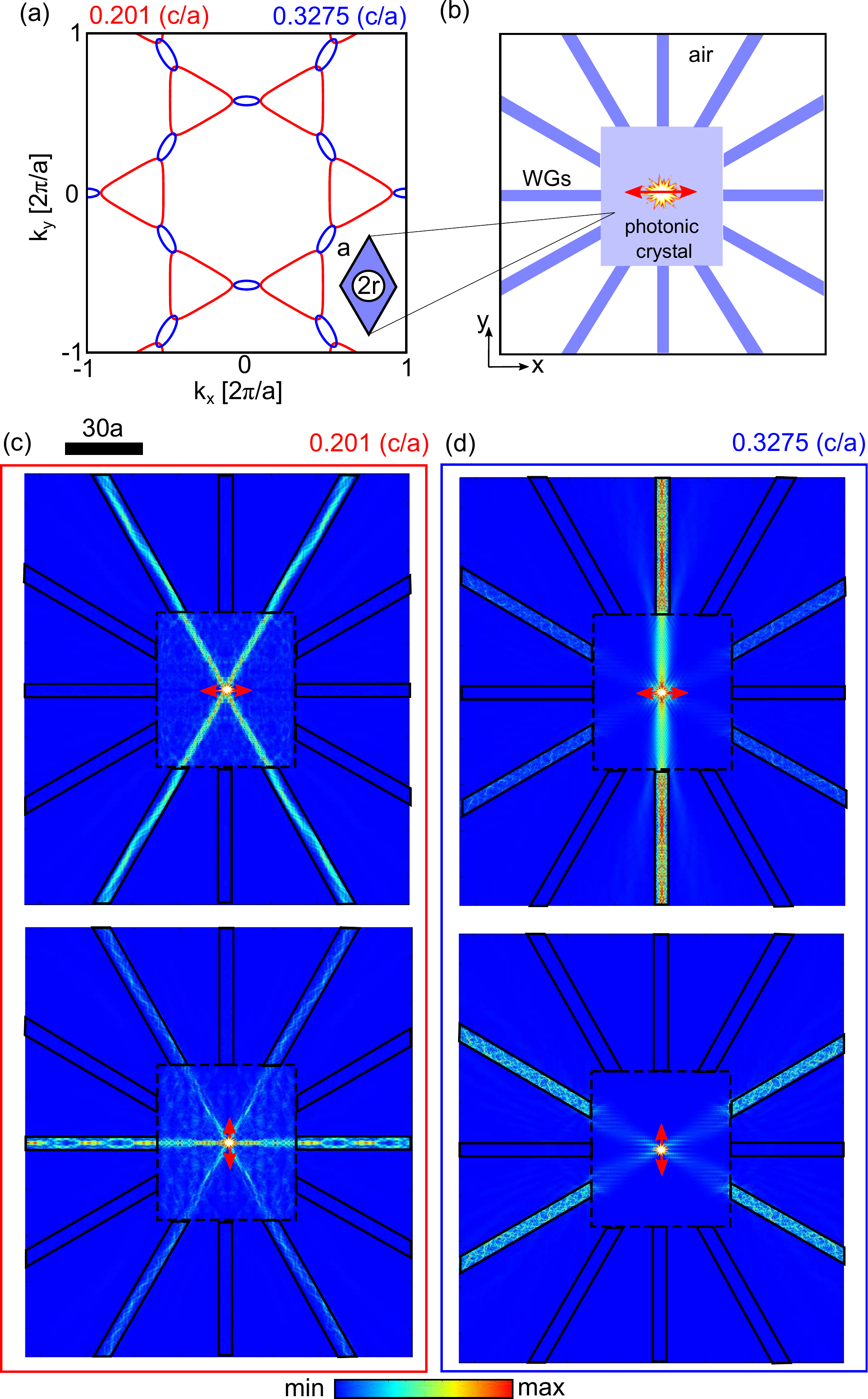}
    \caption{Numerical example of frequency dependent power splitting in a two-dimensional photonic crystal formed by hexagonal air holes of radius $r$ and period $a$ within a high index dielectric (blue region in the (a) inset; relative permittivity: 12). (a) Photonic crystal isofrequency contour for TE polarization at a frequency of 0.201 $c/a$ (red) and 0.3275 $c/a$ (blue), calculated using the plane-wave expansion method~\cite{minkov2020inverse} which qualitatively displays features akin to those of the photonic crystal slab presented in the main manuscript, see for example Fig.~3(c),(f) in the main manuscript. (b) Schematic of the device considered, formed by the 2D photonic crystal (light blue region) connected to twelve surrounding dielectric slab waveguides (dark blue region). An electric dipole is placed in the middle of the photonic crystal.  (c) Poynting vector magnitude at a frequency of  0.201 $c/a$ and (d)  0.3275 $c/a$, for $x$-polarized (top row) and $y$-polarized (bottom row) electric dipole placed in the middle. Note the frequency- and dipole orientation-dependent splitting into the surrounding waveguides as a result of the underlying photonic crystal bandstructure. }
    \label{figS3}
\end{figure}

\end{document}